\documentclass[a4paper,fleqn,usenatbib]{mnras}

\usepackage{mathptmx}

\usepackage[T1]{fontenc}
\usepackage{ae,aecompl}

\usepackage{graphicx}	
\usepackage{amsmath}	
\usepackage{amssymb}	
\usepackage{gensymb}	
\usepackage{natbib} 

\title[Analysis of polar faculae]{Analysis of spatially deconvolved polar faculae}

\author[C. Quintero Noda et al.]{C. Quintero Noda,$^{1}$\thanks{E-mail: carlos@solar.isas.jaxa.jp}
Y. Suematsu,$^{2}$
B. Ruiz Cobo,$^{3,4}$
T. Shimizu,$^{1}$
\newauthor
A. Asensio Ramos,$^{3,4}$
\\
$^{1}$Institute of Space and Astronautical Science, Japan Aerospace Exploration Agency, Sagamihara, Kanagawa 252-5210, Japan\\
$^{2}$National Astronomical Observatory of Japan, 2-21-1 Osawa, Mitaka, Tokyo 181-8588, Japan\\
$^{3}$Instituto de Astrof\'isica de Canarias, E-38200, La Laguna, Tenerife, Spain.\\
$^{4}$Departamento de Astrof\'isica, Univ. de La Laguna, La Laguna, Tenerife, E-38205, Spain
}

\date{Accepted XXX. Received YYY; in original form ZZZ}

\pubyear{2015}

\begin{document}
\label{firstpage}
\pagerange{\pageref{firstpage}--\pageref{lastpage}}
\maketitle

\begin{abstract}
Polar faculae are bright features that can be detected in solar limb observations and they are related to magnetic field concentrations. Although there is a large number of works studying them, some questions about their nature as their magnetic properties at different heights are still open. Thus, we aim to improve the understanding of solar polar faculae. In that sense, we infer the vertical stratification of the temperature, gas pressure, line of sight velocity and magnetic field vector of polar faculae regions. We performed inversions of the Stokes profiles observed with Hinode/SP after removing the stray light contamination produced by the spatial point spread function of the telescope. Moreover, after solving the azimuth ambiguity, we transform the magnetic field vector to local solar coordinates. The obtained results reveal that the polar faculae are constituted by hot plasma with low line of sight velocities and single polarity magnetic fields in the kilogauss range that are nearly perpendicular to the solar surface. We also found that the spatial location of these magnetic fields is slightly shifted respect to the continuum observations towards the disc centre. We believe that this is due to the hot wall effect that allows detecting photons that come from deeper layers located closer to the solar limb.
\end{abstract}




\section{Introduction}

Solar observations of polar regions reveal the existence of small scale bright features called polar faculae. These bright features are detected at high latitudes on the solar disk, above $70$~degrees \citep{Okunev2004,BlancoRodriguez2007}, with white light observations as well as measuring chromospheric spectral lines. Their presence is directly related to the solar inner magnetism, as they follow an 11-year cycle that is anticorrelated to the solar sunspot cycle \citep{Makarov1989}. As the sunspot activity increases, polar faculae regions start to migrate towards the poles until they totally vanish when the maximum of sunspot activity is reached \citep{Makarov2003}. Regarding  their magnetic properties, \cite{Homann1997,Okunev2004,Tsuneta2008b} found that the polar faculae harbour magnetic field strengths in the kilogauss range and that they are unipolar, with the same polarity of the observed global polar field. Later, \cite{Kaithakkal2013} used inversions of spectropolarimetric data to reveal that polar magnetic patches have substructure, with one or more small faculae embedded in larger patches. Additionally, they pointed out that their results are consistent with the hot-wall model \citep{Spruit1976}, which attributes the enhanced brightness of faculae to a depression in the visible surface caused by magnetic pressure, allowing an enhanced view of the hot wall of the flux tube at oblique angles (see \cite{Stein2012} or \cite{Borrero2015} for recent reviews). Although previous studies have provided a deep insight of the polar faculae, there are still some questions unanswered as: which is the vertical stratification of their physical parameters or whether there is any difference in their magnetic configuration when they are observed at different heliocentric angles. We aim to address these questions in this work using the results of two complementary techniques: the spatial deconvolution and the inversion of spectropolarimetric data. The first technique allows us to remove the stray light contamination produced by the spatial point spread (PSF) function of the telescope. The second one let us to infer the physical information of the atmospheric parameters at different heights. Thus, these techniques together, allow us to infer the atmospheric information in faculae regions with no stray light contamination. We aim to improve our understanding of these magnetic structures providing, at the same time, information to develop more accurate theoretical models.

\section{Data analysis}


We analysed the magnetic properties of the polar faculae using observations taken by the Spectropolarimeter ($SP$) \citep{Lites2013} on board $Hinode/SOT$ \citep{Kosugi2007,Tsuneta2008,Suematsu2008,Shimizu2008SOT} on 2007 September 6th between 00:28-05:57 UT. $Hinode/SP$ observed the north solar pole recording the Stokes $I$, $Q$, $U$, and $V$ profiles for the Fe~{\sc i} 6301.5 and 6302.5~\AA \ spectral lines. It used a spectral sampling of 21.55~m\AA, and a pixel size and scanning step of about 0.16 arcsec providing a spatial resolution of 0.32 arcsec. The integration time was 9.6 s per slit position what produces a noise value for the magnetic Stokes parameters of approximately $1\times10^{-3}$ of the continuum signal ($I_c$) \citep[for more information, see][]{Ichimoto2008,Lites2013SP_Prep}.

\begin{figure}
\hspace{+0.2cm}
\includegraphics[width=10.0cm]{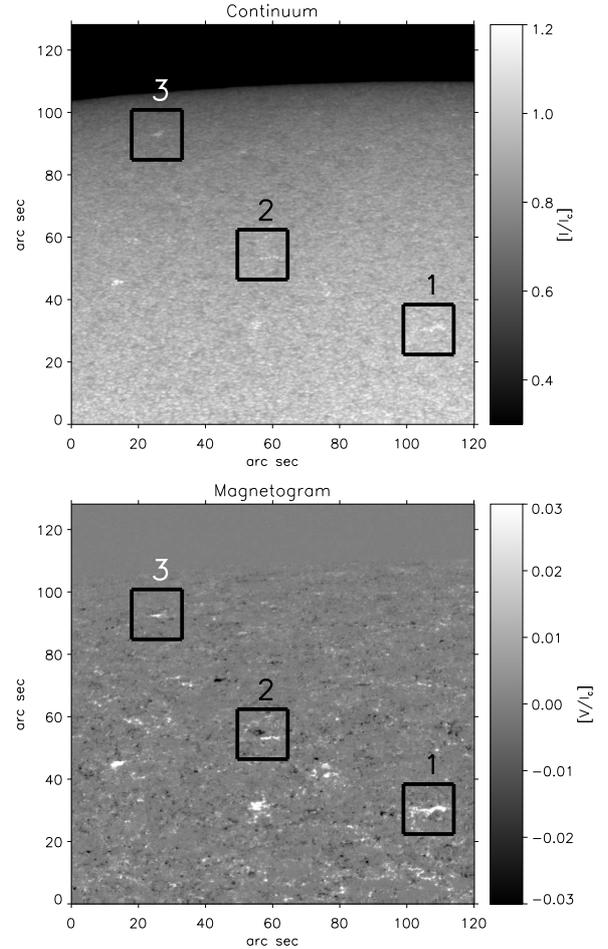}
\caption{Observed map used in this work. Continuum signal (top) and magnetogram (bottom). Black boxes mark the location of the magnetic patches to be examined in detail in following sections. Solar north pole corresponds to the upper part of the field of view.}
\label{cont}
\end{figure} 

Figure \ref{cont} shows in the top panel the continuum signal of the observed map used in this work, the same map used in \cite{QuinteroNoda2016}. We examined a small part of the observed field of view, leaving the off-limb upper region outside the analysis. Bottom row shows a magnetogram, built as the difference between the Fe~{\sc i} 6302.5~\AA \ Stokes~$V$ intensity at $\pm$100~m\AA \ from the line centre. We can see that there is a large number of bright regions in the continuum map that correspond to strong signals in the magnetogram. We marked with numbered boxes the location of the magnetic patches we are going to study in detail later. We selected these patches to cover different heliocentric angles aiming to understand if there is any difference between their physical properties. These three patches, i.e. the centre of the boxes 1, 2, and 3 displayed in Figure \ref{cont}, are located at $\mu$=0.40, 0.32, and 0.20, respectively.

\begin{figure*}
\hspace{-0.3cm}
\includegraphics[width=16.0cm]{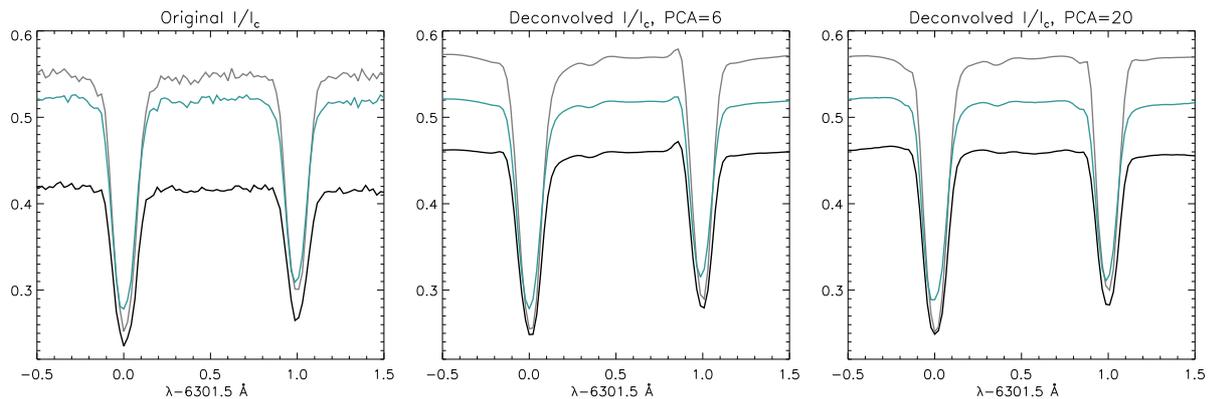}
\caption{Intensity profiles from three selected pixels (different colours) that are located close to the solar limb ($\mu\sim0.15$). First column shows the original profiles. Second and third column display the corresponding deconvolved profiles using 6 and 20 PCA eigenvectors for the reconstruction process.}
\label{artefacts}
\end{figure*} 

\subsection{Deconvolution method}

We have performed the spatial deconvolution of the observed data using the same method presented in \cite{QuinteroNoda2015}. The purpose of this technique is to remove the blurring effects induced by the spatial PSF of the telescope. Traditional deconvolution codes make the deconvolution wavelength by wavelength. The novelty of our code is that it uses a regularization based on a Karhunen-Lo\`{e}ve transformation or principal component analysis (PCA) \citep{Loeve1955} to develop a set of eigenvectors that are appropriate to reproduce the spectral information of each pixel. At the same time, the polarimetric signal information is included in the first eigenvectors while the rest of eigenvectors encompasses the information related to the noise. Thus,
we can reduce the noise by selecting just a few eigenvectors. Additionally, a small number of eigenvectors leads to a less time-consuming computational process. After selecting the PCA eigenvectors, the code employs a Richardson-Lucy \citep{Richardson1972,Lucy1974} algorithm to perform subsequent iterations until the stray light contamination introduced by the spatial PSF is removed from the data (see \cite{QuinteroNoda2015} for more information). 

We have found in previous works that a small number of PCA eigenvectors is more than enough to reproduce the Stokes profiles without introducing significant differences \citep[see][]{RuizCobo2013,QuinteroNoda2015}. We followed in this work the same procedure but we detected some artefacts in the Stokes~$I$ profiles of pixels located close to the limb (at $\mu\sim0.15$), see Figure \ref{artefacts}. If we examine the Stokes profiles of the second column, we will find a positive lobe on the red wing of the Fe~{\sc i} 6302.5~\AA \ line. This positive lobe, above the continuum value, indicates the presence of a satellite line in emission. Something difficult to justify at this wavelength position because the core of the line is in absorption. Therefore, as these profiles are close to the solar limb, we believe that this is an effect of overcompensation. To check if this assumption is correct, we first changed the number of iterations used by the deconvolution algorithm, and then the number of PCA eigenvectors used for the reconstruction of the Stokes profiles. The first study revealed no clues, i.e. a large number of iterations could generate artefacts but those ones were already present since the second iteration. We proceeded, then, to increase up to 20 the number of PCA eigenvectors used for the reconstruction of the Stokes profiles. If we look at the rightmost panel, we can see that the artefacts have disappeared. We believe that the reason behind these artefacts is the large variety of Stokes $I$ profiles in the pixels close to the limb. In that sense, we can find absorption profiles for on disc pixels while the Stokes $I$ profiles appear in emission for off disc pixels \citep[for instance, ][]{Lites2010}. Therefore, the shape of the Stokes $I$ profiles of this map is the most complicated and varied we have ever studied.

We decided to use a large number of PCA eigenvectors for the reconstruction process to avoid the appearance of any artefact. The final number of eigenvector is (20, 7, 5, 10) for ($I$, $Q$, $U$, $V$). It is worth to note that the number of PCA families for the linear polarization Stokes parameters has increased with respect to the ones used in the previously cited works. This is because the linear polarization signals become more important in the polar faculae and can also show a complex shape in some pixels. We performed 7 iteration steps in the deconvolution process improving the original continuum contrast from 4.5 to 7 per cent. We stopped the iteration process in this step because it provides an increase factor similar to the one obtain in previous works, e.g. \citep{QuinteroNoda2015} used disc centre observations with low noise, improving the continuum contrast a factor of 1.57 (from 7.6 to 11.9 per cent). In addition, we also computed the relation factor between the disc centre contrast obtained in the previously mentioned work and the one inferred here for limb observations. We obtained a value of $f=7/11.9=0.59 $, that is comparable with the one estimated by \cite{SanchezCuberes2000} for $\mu=0.3$ (see Fig. 3 of that work). 

Finally, we would like to mention some drawbacks of this method. One is that it is an iterative process based on a maximum-likelihood approach and, therefore, is sensitive to overfitting. This means that we have to check that the number of iterations is large enough to improve the image quality but small enough to avoid introducing artefacts. An additional limitation is that the set of PCA eigenvectors we use to reconstruct the deconvolved profiles is created from the observed data. This is because we do not know which were the Stokes profiles before they were affected by the spatial PSF of the telescope.

\subsection{Deconvolution results}

\begin{figure}
\hspace{-0.3cm}
\includegraphics[width=11.3cm]{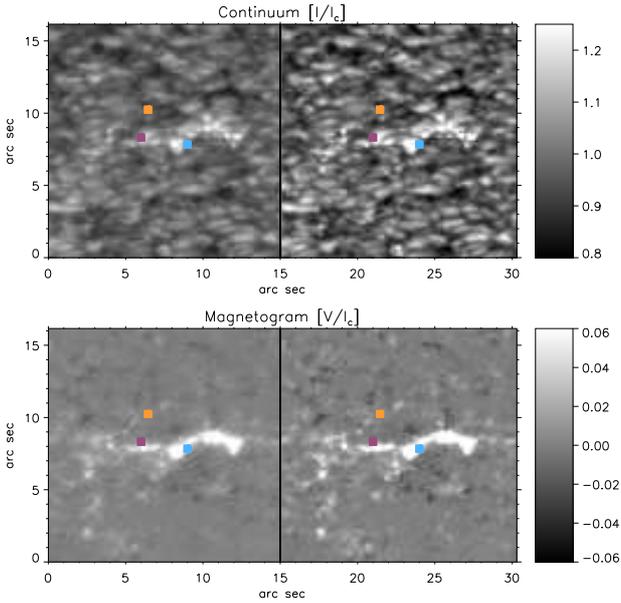}
\caption{Comparison between original (left column) and deconvolved (right column) data. Top panel shows the continuum intensity while bottom panel displays a magnetogram. Coloured squares indicate the position of the pixels shown on Figure \ref{compar3}.}
\label{compar2}
\end{figure}

Figure \ref{compar2} shows the comparison between the original continuum and magnetogram (left column) and the corresponding deconvolved quantities (right column) for a small fragment of the map. We reduced the field of view to make easier the detection of the differences induced by the deconvolution process. The continuum image looks sharper due to the increase of the continuum contrast. The same improvement is found for the magnetogram map where the deconvolved magnetic structures display larger intensity values. In addition, we are also able to detect some structures that appear blurred in the original data. All of these results are in agreement with \cite{QuinteroNoda2015}. 

\begin{figure*}
\hspace{-0.3cm}
\includegraphics[width=16.0cm]{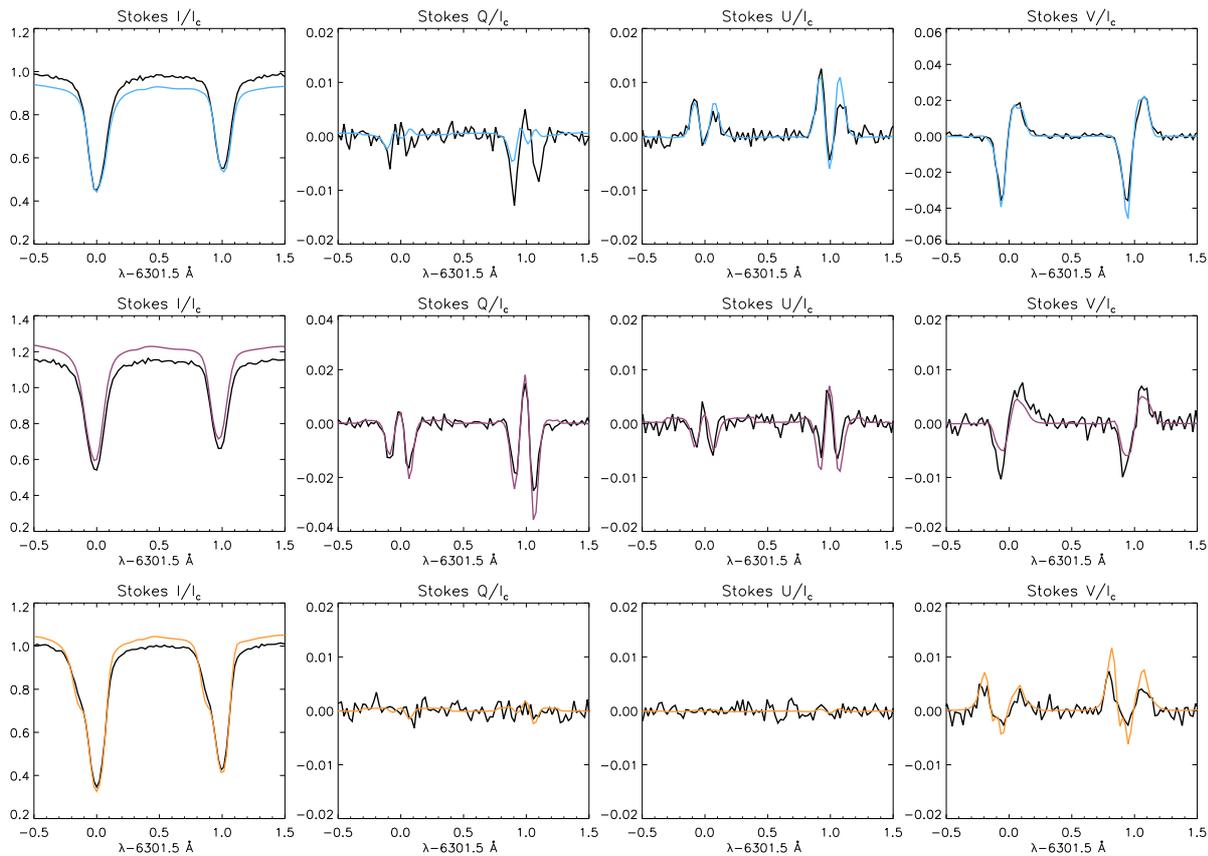}
\caption{Comparison between original (black) and deconvolved (coloured) profiles. We indicate with coloured squares their location in Figure \ref{compar2}.}
\label{compar3}
\end{figure*}

If we examine Figure \ref{compar3}, we can see that the deconvolution method produces several changes in the Stokes profiles. Regarding Stokes $I$, continuum intensity values can be larger or lower than the original ones, directly related to the increase of the continuum contrast. In addition, some spectral features, as satellite lobes \citep{BellotRubio2009,QuinteroNoda2016} are enhanced in the deconvolved data (see orange Stokes $I$ in Figure \ref{compar3}). Concerning the polarimetric parameters, we can see a similar behaviour, i.e. deconvolved profiles can show larger or lower intensities. However, it seems that the deconvolution process does not abruptly change their shape, except the blue coloured Stokes~$Q$ profile. Therefore, we do not expect significant changes in the area/amplitude asymmetry of the deconvolved profiles \citep{QuinteroNoda2015}.

\begin{figure*}
\hspace{-0.3cm}
\includegraphics[width=17.0cm]{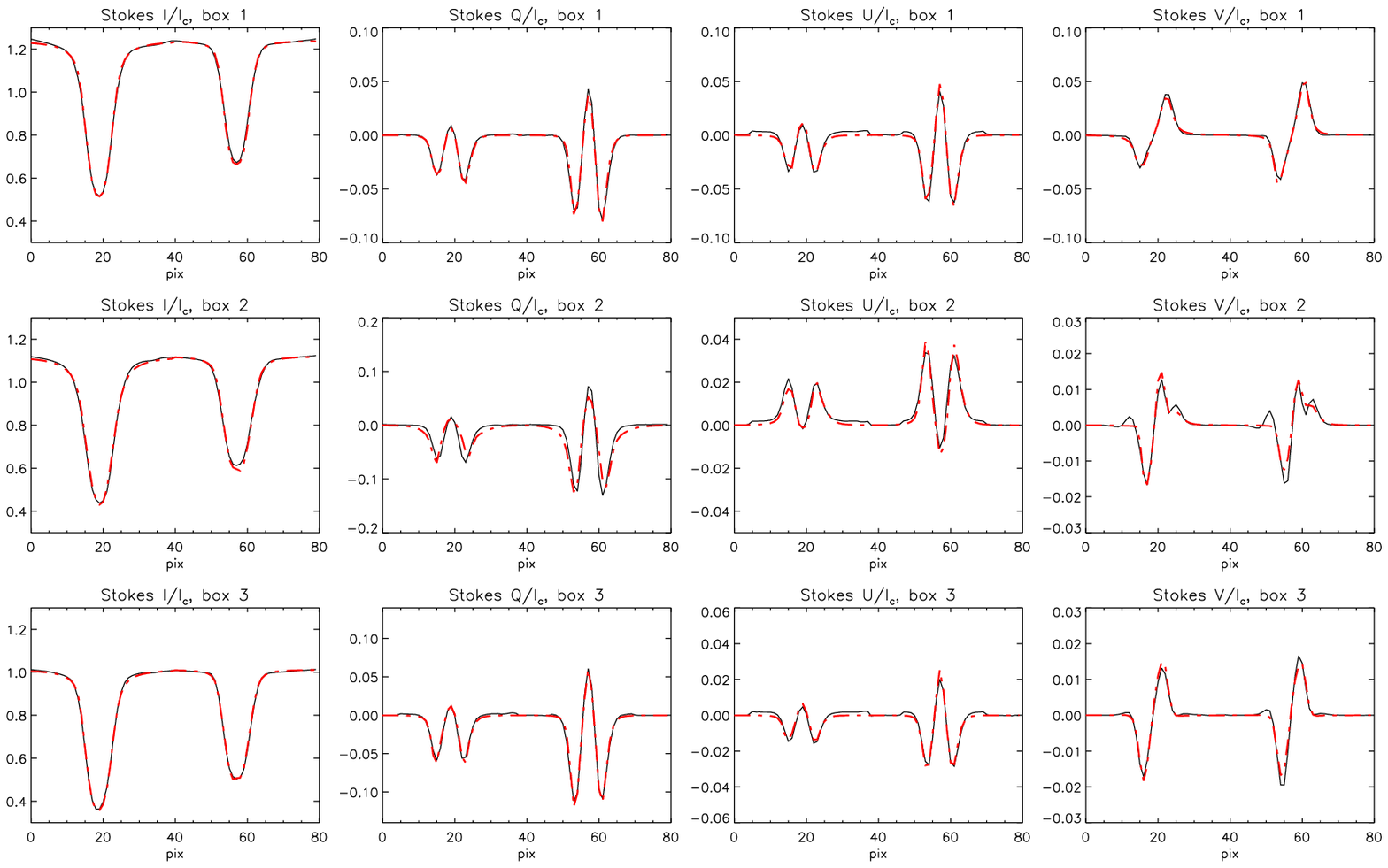}
\caption{Results from the inversion of Stokes profiles located at different heliocentric values. Each individual profiles belongs to the magnetic patch centred in the boxes shown in Figure \ref{cont}. Original profiles are plotted in black while the results from the inversion are displayed in red dashed lines.}
\label{fits}
\end{figure*}

\section{Inversion of Stokes profiles}

We have inferred the atmospheric parameters of the polar faculae inverting the Stokes profiles from 3 different regions (see boxes in Figure \ref{cont}). These three regions are located at the heliocentric angles $\mu$=0.40, 0.32, and 0.20. We carried out the inversion of the Stokes profiles using the {\sc sir} \citep[Stokes Inversion based on Response functions;][]{RuizCobo1992} code, which allows us to infer the atmospheric parameters as a function of the continuum optical depth at each pixel independently.

\subsection{Configuration}

As the deconvolution process removes the stray light contribution of the spatial PSF, we do not need to include a stray light component in the inversion process. Consequently, we used a single atmospheric component to reproduce the observed profiles. The selection of nodes for each quantity is done by an automatic algorithm. This algorithm chooses how many nodes we use based on the number of zeros of the response function. Although this algorithm has been available long time in {\sc sir} as an additional option, it has never been described in detail. Therefore, we plan to explain here how it works. For simplicity, we only focus on the intensity profile although the explanation is applicable to any of the four components of the Stokes vector. The inversion is carried out by modifying the values of the atmospheric parameters until we find a minimum of the merit function:

\begin{equation}
\chi^2 = \sum_{i=1}^4 \sum_{j=1}^N w_{ij} \left[ I_i^\mathrm{obs}(\lambda_j) - 
I_i^\mathrm{syn}(\lambda_j) \right]^2,
\end{equation}
where $w_{ij}$ are a set of weights that are equal to the inverse of the noise variance. This is for the standard case although it can be modified at will.  $I_i^\mathrm{obs}(\lambda_j)$ and $I_i^\mathrm{syn}(\lambda_j)$ are the observed and synthetic Stokes profiles ($i=1,2,3,4$ for Stokes $I$, $Q$, $U$ and $V$, respectively) at a wavelength $\lambda_j$. Lastly, $N$ is the number of wavelength points. 

The optimization of the $\chi^2$ is done using a Levenberg-Marquardt algorithm, which makes use of the derivatives of the merit function with respect to the physical conditions at the nodes. For instance, 
\begin{equation}
\frac{\partial \chi^2}{\partial T(\tau_k)} = \sum_{i=1}^4 \sum_{j=1}^N 2 w_{ij} 
\left[ 
I_i^\mathrm{obs}(\lambda_j) - I_i^\mathrm{syn}(\lambda_j) \right] 
R_T(\lambda_j,\tau_k),
\end{equation}
where $R_T(\lambda_j,\tau_k)$ is the response function of the temperature at optical depth $\tau_k$ (where $\tau$ refers to the continuum optical depth) for the wavelength point $\lambda_j$ \citep{Landi1977}. If we assume, for instance, that in the current atmospheric model, we have $I^\mathrm{obs} - I^\mathrm{syn}>0$ for all wavelengths. Thus, we would need to increase $I^\mathrm{syn}$ to reach a better fit. If the response function $R_T(\lambda_i,\tau_j)$ is positive below $\tau_j$, and negative over this optical depth, an optimal solution can be obtained by a positive perturbation of the temperature below $\tau_j$ and a negative one over it. Therefore, we would need two nodes for the temperature stratification in the iterative process. This empirical idea  establishes that the number of nodes in a given physical quantity is equivalent to the number of zeros of the corresponding derivative of $\chi^2$. 

In our case, we limited the maximum number of nodes available aiming to avoid too complex solutions. We take advantage of the knowledge acquired in \cite{QuinteroNoda2014b} and \cite{QuinteroNoda2015} to determine this maximum number of nodes. We allowed seven nodes for the temperature T($\tau$), five for the line of sight (LOS) component of the velocity v$_{\rm los}$($\tau$), five for the magnetic field strength B($\tau$), three for the inclination of the magnetic field $\gamma$($\tau$), and two for the azimuthal angle of the magnetic field $\phi$($\tau$). The automatic algorithm computes the optimum number of nodes for each atmospheric parameter at each iterative step. This optimum number of nodes is always lower or equal to the maximum number of nodes we established before.

Moreover, contrary to what happens with on disc observations, the line of sight component is not perpendicular to the solar surface. This condition introduces an additional complexity on the observed velocity stratification and, for this reason, we consider justified inverting the microturbulence with one node. On the other hand, the macroturbulence is null and not inverted. Since each node corresponds to a free parameter, our model could include up to 23 free parameters in case the automatic algorithm selects the maximum number of nodes allowed. Although this number is relatively large, we want to note that we observe complex Stokes profiles that are also affected by projection effects. Additionally, we aim to study the height stratification of the atmospheric parameters, what demands a good fit of the Stokes profiles. 

During the inversion process, we compare the observed profiles and the synthetic ones at each iteration. This comparison is done after convolving the synthetic profiles with the spectral transmission profile of $Hinode/SP$ \citep{Lites2013}. Additionally, as the inferred physical parameters are reliant on the initial conditions, we minimize this effect by inverting each individual pixel with ten different initial atmospheric models. These initial random atmospheric models were created as in \cite{QuinteroNoda2015}, i.e. we linearly modified the HSRA model \citep{Gingerich1971} with random perturbations. Finally, from the available solutions, we pick the one that produces a smaller $\chi^2$ value.

\begin{figure*}
\includegraphics[width=16.5cm]{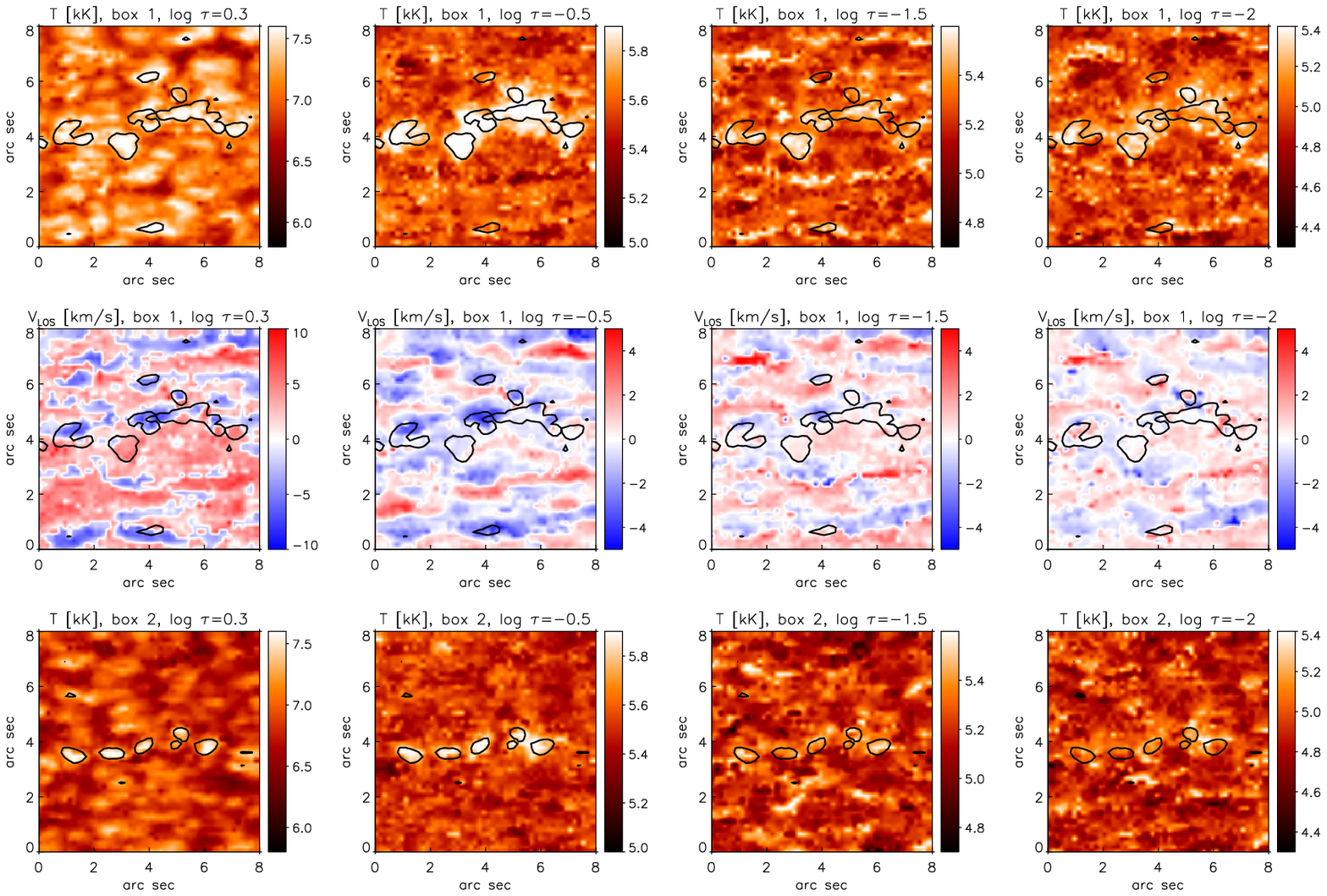}
\includegraphics[width=16.5cm]{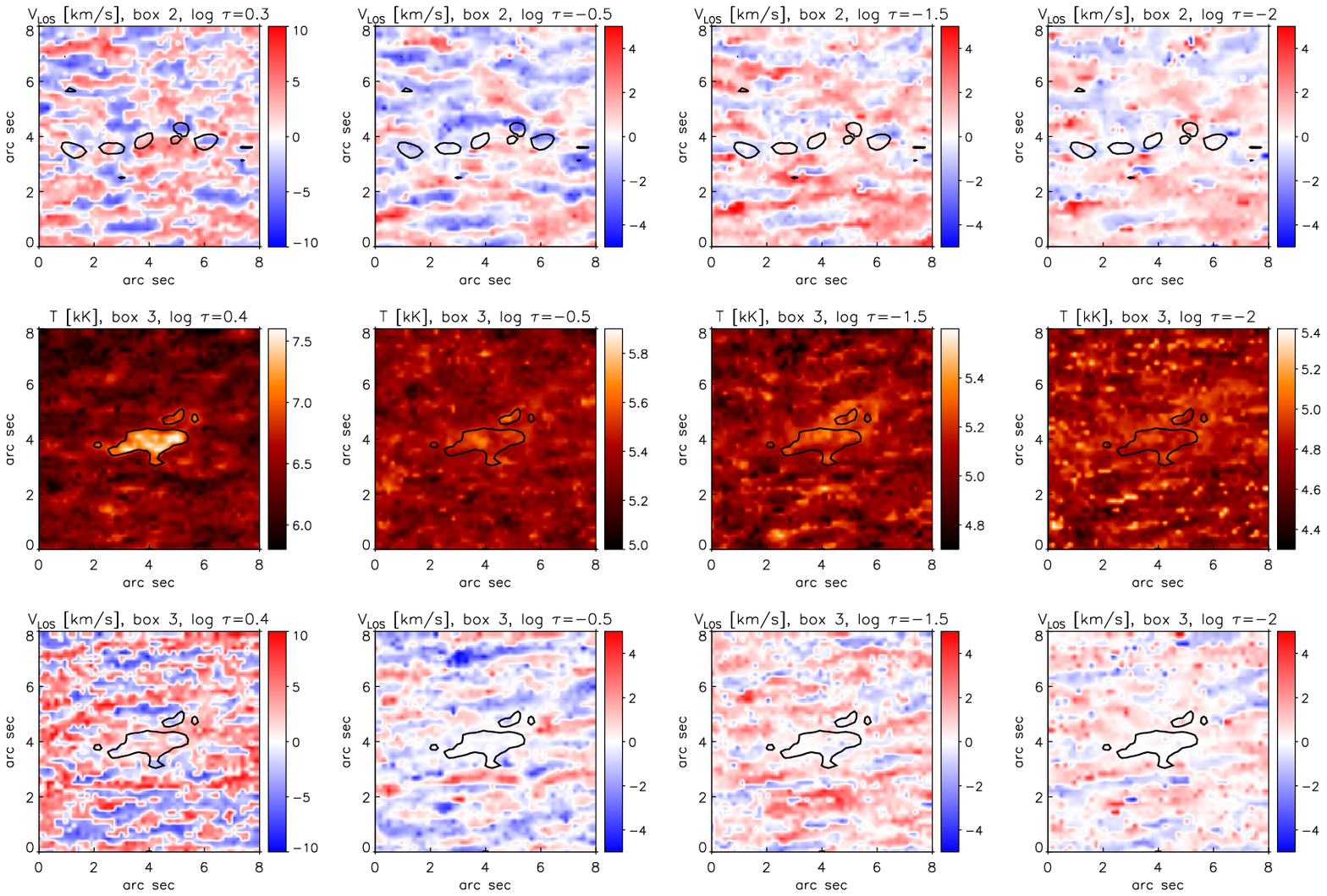}
\caption{Temperature (odd rows) and LOS velocity (even rows) spatial distribution at different atmospheric heights (different columns). Each panel corresponds to the boxes marked in Figure \ref{cont} although we displayed a smaller field of view centred in the magnetic patch. Blue colour in the LOS velocity maps indicate that the material is moving towards the observer while red colours designate the opposite direction. Contours correspond to the location of the faculae in the continuum intensity map.}
\label{2d}
\end{figure*}

\begin{figure*}
\includegraphics[width=16.5cm]{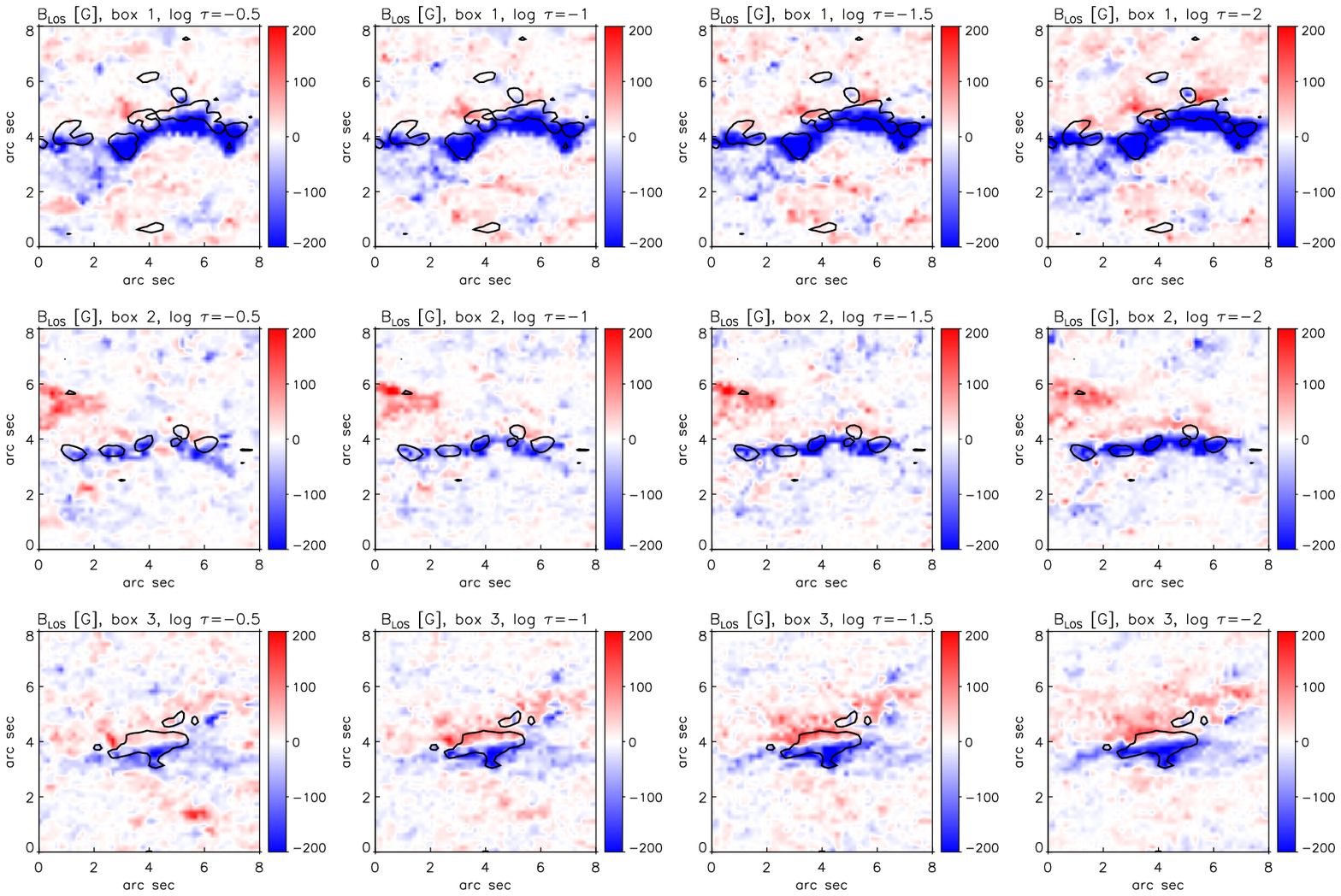}
\caption{Same as Figure \ref{2d} but for the magnetic field along the line of sight. White colour indicates a null magnetic field strength or an inclination value of 90 degrees. Blue displays the regions where the magnetic field inclination is larger than 90 degrees while red depicts the regions where the magnetic field inclination is lower than 90 degrees. Contours correspond to the location of the faculae in the continuum intensity map.}
\label{2dfield}
\end{figure*}

\subsection{Selected Stokes profiles}

We show in Figure \ref{fits} some examples that illustrate the accuracy of the inversions. Each row shows the four Stokes parameters (see different columns) from a given pixel. These pixels, from top to bottom, belong to the magnetic patches located in boxes number one, two and three, respectively (see Figure \ref{cont}).

The first thing that draws our attention is the large linear polarization signals. In fact, the Stokes $Q$ amplitude is always larger than the Stokes $V$ maximum signal, something that is very infrequent in quiet Sun disc centre observations.  We can also see that these linear polarization profiles are fitted with high accuracy. Regarding the Stokes $V$ profiles, we also obtained very good fits for the profiles of top and bottom rows. However, the fit of the Stokes $V$ parameter shown in the middle row is less accurate. This is because the profile displays multiple lobes. On the red wing, we can see a bump, that could be related to abrupt gradients along the LOS \citep[see, for instance,][]{Shimizu2008,QuinteroNoda2014}. On the other hand, we can see a positive satellite lobe at a bluer wavelength position than the blue lobe, indicating the presence of unresolved magnetic structures inside the observed pixel. Thus, we believe that the only way to properly solve this type of profiles is using a combination of two magnetic components. Fortunately, these profiles are less common than the ones shown in the top and bottom rows. Finally, in order to quantify the goodness of these fits, we computed the reduced $\chi^2$ following \cite{QuinteroNoda2015} (see Equation~11 of that work). We obtained that ${\chi^2}=[3.99,5.36,3.05]$ for the three set of Stokes profiles shown in Figure~\ref{fits}, from top to bottom row, respectively. As we expected, the Stokes profiles of the middle panel are the worst fitted while the others are better reproduced.

\begin{figure*}
\hspace{+0.3cm}
\includegraphics[width=15.0cm]{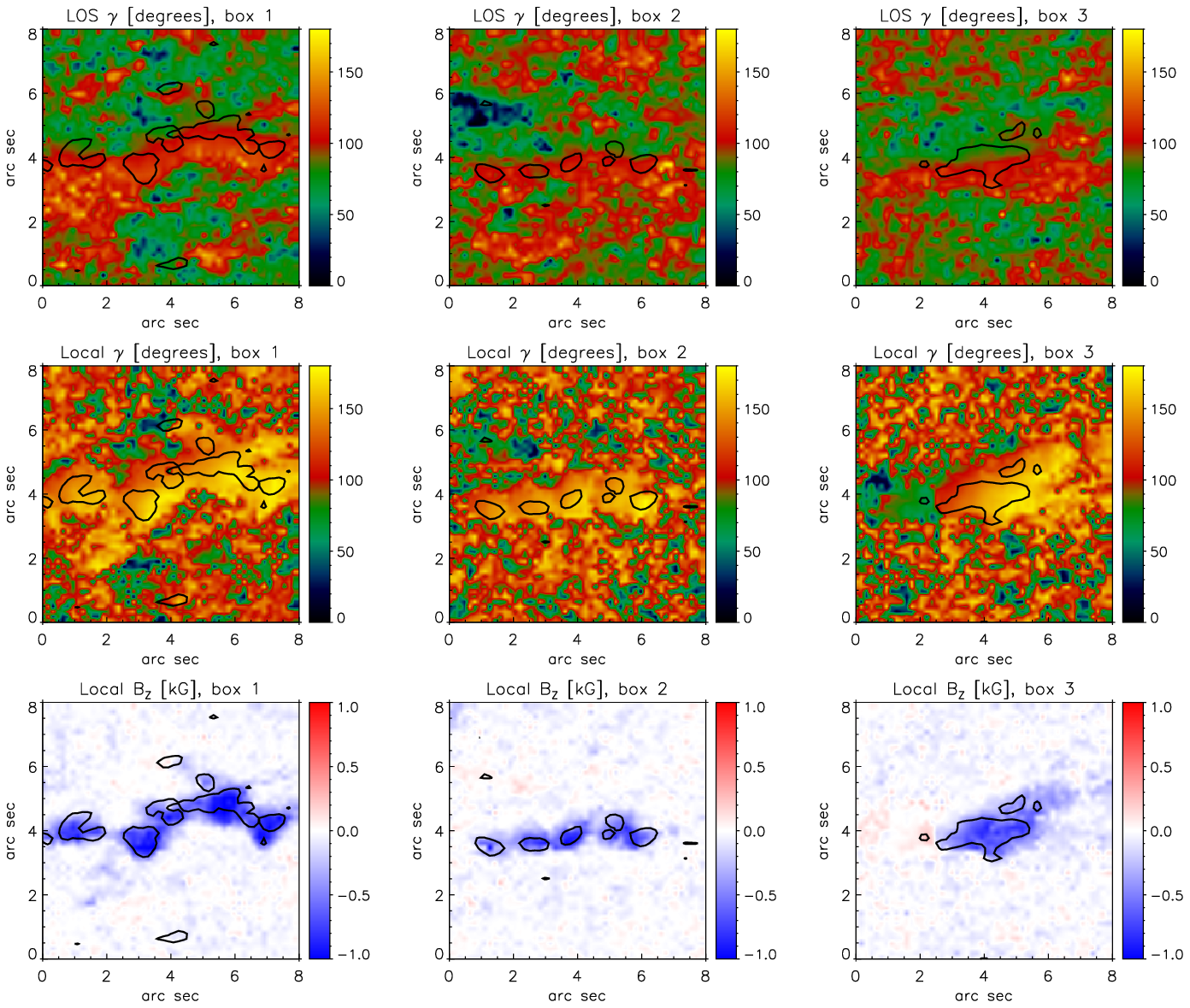}
\caption{Comparison between LOS magnetic field inclination (first row) and local magnetic field inclination (second row) for the three different patches studied in previous sections, different columns. In the first row, an inclination of zero degrees means a magnetic field pointing to the observer while ninety degrees indicates a magnetic field perpendicular to the line of sight. For the second row, zero degrees signifies that the magnetic field is perpendicular to the solar surface while ninety degrees means that the magnetic field is parallel to the solar surface.  Bottom row shows the results for the vertical component of the magnetic field at local coordinates. These results should be compared with the third column of Figure \ref{2dfield}. All the values represented in the panels correspond to the optical depth $\log$ $\tau=-1.5$. Contours correspond to the location of the faculae in the continuum intensity map.}
\label{local}
\end{figure*}

\subsection{Spatial properties}

We show in Figure \ref{2d} the spatial distribution of the temperature and the LOS velocity at different heights. We depict the results for the three magnetic patches marked with boxes in Fig. \ref{cont}. We also marked in every panel the location of the polar faculae in the continuum intensity map (see black contours). 

If we start with the temperature of the first case, top leftmost panel, we can see that the granulation pattern is replicated by the temperature at the optical depth $\log$~$\tau=0.3$. This optical depth is deeper than the usual value at disc centre observations, i.e. $\log$ $\tau=0$. This shift of optical depths is produced because the LOS is not perpendicular to the solar surface,  and, consequently, the detected photons come from higher layers in the photosphere. This means that limb observations provide information of slightly higher geometrical heights for the same optical depth. Concerning the temperature stratification of the polar faculae we can see that there is a hot region that corresponds to the bright patch seen in the continuum intensity map (black contours). This enhanced region is also present at higher optical depths, see following columns of the same row, being always hotter than its surroundings. Moving on to the LOS velocity, second row, we can see that the polar faculae displays a mixed of blue- and redshifted velocities at lower heights that quickly decrease to almost zero at higher heights. 

The second case, box number two in Figure \ref{cont}, displays a similar behaviour, see third and fourth rows. There is a hot region, corresponding to the polar faculae, visible at all layers. This magnetic patch also shows blue- and redshifted velocities at the bottom of the photosphere and almost null values at higher layers. 

The third case, box number three in Figure \ref{cont}, shows small differences with respect to the previous cases, see fifth and sixth rows. The granulation pattern is very poor because the heliocentric angle is large, $\mu \sim 0.2$. In this case, the detected photons come from even higher geometrical heights. This produces that we cannot recover the granulation pattern even at the lowest optical depth where we can retrieve reliable values (approximately $\log$ $\tau=0.4$). This means that the shift in geometrical heights is too high to show temperatures that correlate with the continuum intensity map. However, if we examine the temperature distribution inside the facular region we find a hot patch, as before, that is present for all heights, although it is less clear at higher layers (see rightmost column). Lastly, the LOS velocity has a similar behaviour with a mixture of pixels with blue- and redshifted velocities at the bottom of the atmosphere while the velocity pattern is clearer at higher heights showing null velocities. 

The inversion result for the LOS magnetic field, i.e. the magnetic field strength times the cosine of the magnetic field inclination, is displayed in Figure \ref{2dfield}. As before, we show in different rows the results for the polar faculae enclosed by the boxes shown in Figure \ref{cont}. Looking at the first row, the region at higher $\mu$ value, we can see that a single polarity is predominant (blue colour). The same behaviour can be found for the rest of the cases. In addition, if we look to the first case at higher heights, rightmost column, an opposite component (red colour) can be detected at the north part (closer to the solar limb) of the magnetic structure. Moreover, this opposite component is more prominent as we analyse regions with lower $\mu$ values, middle and bottom rows, indicating a connection with projection effects. 

Remarkably, we also found that the location of the magnetic field patches is shifted southward respect to the position of the polar faculae on the continuum intensity map (black contours). We believe that this could be due to the hot wall effect in these regions. This effect produces that we detect the continuum intensity, that comes from lower heights, closer to the solar pole. Moreover, the case showing a largest shift (upper panel) is the one with highest magnetic field strength. Finally, we detected that the spatial size of the polar faculae is increasing with height in all the studied physical quantities, indicating that the magnetic structure expands with height. This result is in agreement with \cite{Tsuneta2008b} and \cite{Ito2010}.

\section{Magnetic fields at local solar coordinates}

\subsection{Method}

There are two questions that arise from the previous results. One of these question is: Is the magnetic field of polar faculae perpendicular to the solar surface? The second one could be: How large is the strength of the vertical component of the magnetic field? However, to answer both questions, we need to calculate the values of the magnetic field inclination at local coordinates. To this purpose, we used the {\sc idl} routine \textit{$r\_frame\_asp.pro$} that can be found in the Advance Stokes Polarimeter \citep[ASP, ][]{Elmore1992} {\sc idl} libraries. The information required to launch the program was obtained from the header of each slit and from the inversion results showed in previous sections. In addition, the transformation to local coordinates depends on the azimuth value so we need to solve the 180 degrees ambiguity before using the mentioned libraries. Moreover, there are several ways to solve the 180 degrees ambiguity,  see, for instance the review of \cite{Metcalf2006}. In our case, we minimise a merit function built as the sum of two terms. The first one is the squared of the divergence of the magnetic field in the LOS reference system integrated over the full image. The second term of the sum is the squared of the LOS component of $\nabla \times B$.  The reason why we chose minimising these two terms in conjunction is because this seems to be the most promising way to compute the 180 degree ambiguity (see the conclusions of \cite{Metcalf2006}). Lastly, we evaluated the previous quantities at the optical depth where the uncertainty of the inferred magnetic field vector is minimum. In our case, this is approximately at $\log \tau=-1.5$. To obtain the minimum value of the merit function we use a genetic algorithm.

\subsection{Results}

We plot in the first row of Figure \ref{local} the LOS magnetic field inclination results at $\log$ $\tau=-1.5$. Each column corresponds to the inversion results of the three boxes shown in Figure~\ref{cont}. We can see that green and red colours are predominant indicating that the magnetic field inclination of the three selected patches is between $80<\gamma<110$ degrees. However, the results of the magnetic field inclination at local coordinates, second row, show an opposite behaviour. The inclination values in the polar faculae are mainly vertical (with respect to the solar surface) after the coordinate transformation. Note that now the yellow colour is the most abundant in the magnetic regions. Therefore, if we plot the results of the local vertical component of the magnetic field, bottom row, we find a very different picture than the one displayed in the third column of Figure \ref{2dfield}. In this case, the vertical component of the magnetic field exceeds 1 kG at the location of the polar patches. Additionally, although the transformation to local coordinates change the inclination values it does not alter the field strength. Thus, the vertical component of the magnetic field still shows low values outside the polar faculae. This produces a smoother landscape for the vertical component of the magnetic field (third row) than the one obtained for the local magnetic inclination (second row). Moreover, there is no presence of opposite polarities at the limb side (upper part of the panels) of the polar faculae as we found in the LOS reference system (see Figure \ref{2dfield}).

\begin{figure*}
\includegraphics[width=17.0cm]{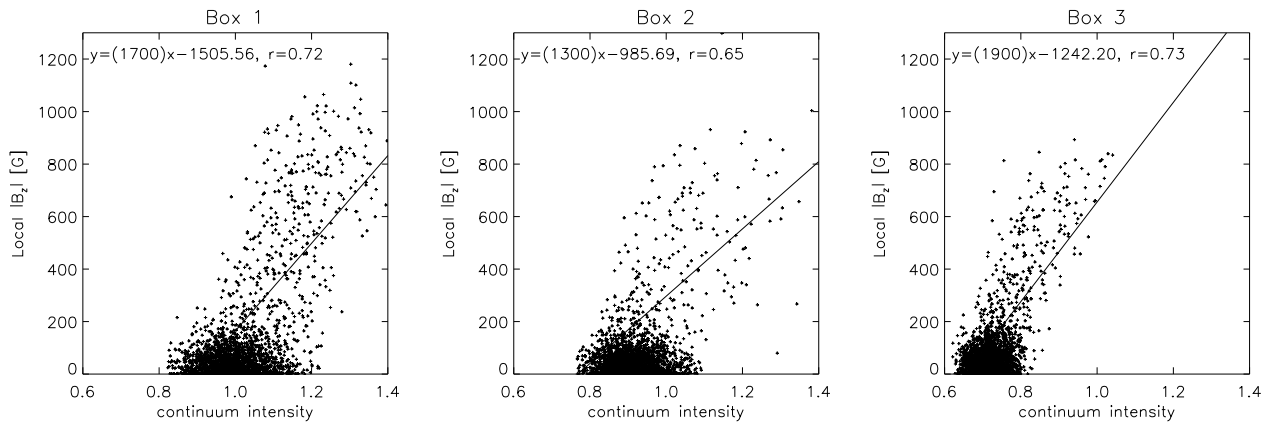}
\caption{Scatter plot of the continuum intensity versus the vertical magnetic field at local coordinates. We compute the magnetic field at $\log \tau=-1.5$ and we displayed its absolute value. We indicate on the top of each panel the results of the linear fit of the displayed values and the correlation between them.}
\label{correl}
\end{figure*}

\subsection{Spatial correlation between the local vertical field and the continuum intensity}

We show in Figure \ref{correl} the correlation between the local vertical field and the continuum intensity for the three examined cases (boxes in Figure \ref{cont}). As mentioned before, the local vertical field is computed at $\log \tau=-1.5$. We can see that there is a linear relation between the field strength and the continuum value, i.e. brighter faculae seem to be associated with stronger magnetic fields. We computed a linear regression using the data displayed in Figure \ref{correl} to estimate the previous assumption. We found that the slope of the linear regression is always positive confirming that higher continuum intensities are related to larger magnetic field strengths. In addition, the correlation between both quantities is high for the three magnetic structures. This is because higher magnetic pressures generate larger depressions in the visible surface. This produces the detection of photons that come from deeper and hotter layers what generates higher continuum intensities. However, we cannot confirm if this correlation is a general property of the polar faculae without performing a wider statistical study.

\section{Discussion}

We would like to start discussing one limitation that we should bear in mind when we analyse the results presented in this work. As we move closer to the limb, the detected photons come from higher layers in the atmosphere and they probably correspond to different spatial locations along the line of sight. Thus, projections effects are large and we cannot truly distinguish what it is the polar structure and what is not. In this regard, we would like to remark that we cannot guarantee that our interpretation is completely correct.

The second aspect we want to discuss is the inferred velocity pattern for the polar faculae. We found that there is a mix of red- and blueshifted velocities at lower heights that decrease to almost zero at higher layers in the atmosphere. We still have no clear explanation for this pattern but we believe that this velocity distribution could be produced by a material that is moving under the gravity as the main drive.

Finally, the last topic we want to discuss is the large difference between the spatial distribution of the LOS magnetic field and the local vertical magnetic field. The first one shows opposite polarities while the second one is unipolar. We believe that this large difference is due to the geometry of the magnetic field lines that conform the flux tube. These field lines are mainly vertical respect to the solar surface at lower layers. However,  they expand with height, being more inclined respect to the solar surface, as a consequence of the decrease of the external gas pressure. Therefore, if we look to these bundle of magnetic field lines with a certain angle, we will see that the closer ones are pointing towards us. On the other hand, the lines that belong to the furthest part of the flux tube expand in the opposite direction, pointing away from us. Thus, the expansion of magnetic field lines explains why we detected opposite polarities for the LOS magnetic field and also explains why this effect is stronger as we move closer to the limb.

\section{Conclusions}

We have boarded for the first time the spatial deconvolution of polar faculae regions. We encountered some problems due to the large variety of shapes the Stokes $I$ profile presents, i.e. from absorption to emission profiles. We solved these problems increasing the number of PCA eigenvectors used in the deconvolution process. Thanks to the spatial deconvolution, we enhanced the continuum contrast, the signal to noise ratio, and the visibility of the magnetic structures. We checked the effect of the deconvolution process in the polarization profiles finding that the method alters the amplitude of the profile but not its spectral shape. Later, we inverted the Stokes parameters pertaining to the polar faculae using a single magnetic component. We allowed  gradients along the line of sight for the atmospheric parameters in order to study their height stratification. In addition, thanks to the spatial deconvolution, we did not need to use a stray light component during the inversion process. This condition simplifies the inversion configuration and provides robustness to the results, making them independent of the chosen stray light profile. The inversion results were very accurate although, in a few cases, the complexity of the Stokes profiles could demand the use of two magnetic components. However, our results seem to be reliable, even for those complex profiles. 

The physical information of the examined magnetic patches can be summarized as hot regions with very low LOS velocities, and highly inclined LOS magnetic fields of moderate strength. However, when we analysed the magnetic fields in the local reference frame, we discovered that the very inclined LOS magnetic fields correspond to vertical magnetic fields (with respect to the solar surface) which strength is above 1~kG. Thus, we can conclude that the polar magnetic patches are localised regions of hot material that harbour a strong vertical magnetic field of single polarity. We also found that the spatial location of this magnetic field is slightly shifted respect to the continuum observations towards the disc centre. We believe that this is due to the hot wall effect that allows detecting photons that come from lower heights located closer to the solar limb.  
 
Regarding the future work, we need to analyse a larger number of magnetic regions using, for example, all the maps that belong to the same observation campaign, see \cite{Kaithakkal2013}. To do that, we plan to take advantage of the deconvolution process and the inversion analysis to properly retrieve the atmospheric information from the Stokes profiles. This study will provide statistical verification of the conclusions presented in this work. Moreover, it is also interesting to extend these studies to other type of faculae, e.g. the equatorial faculae.

\section*{Acknowledgements}
AAR acknowledges financial support through the Ram\'{o}n y Cajal fellowship and the project AYA2014-60476-P (Solar Magnetometry in the Era of Large Solar Telescopes). BRC acknowledges financial support through the Project No. ESP2014-56169-C6-2-R funded by the Spanish Ministry of Economy and Competitiveness. \textit{Hinode} is a Japanese mission developed and launched by ISAS/JAXA, collaborating with NAOJ as a domestic partner, NASA and STFC (UK) as international partners. Scientific operation of the Hinode mission is conducted by the Hinode science team organized at ISAS/JAXA. This team mainly consists of scientists from institutes in the partner countries. Support for the post-launch operation is provided by JAXA and NAOJ (Japan), STFC (U.K.), NASA, ESA, and NSC (Norway).

\bibliographystyle{mnras} 
\bibliography{polar} 

\bsp	
\label{lastpage}
\end{document}